\documentclass[prb,twocolumn,showpacs]{revtex4}
\usepackage{amsmath}
\usepackage{dcolumn}
\usepackage{graphicx}

\begin{document}

\title[Short title for running header]{Why the anti-nodal quasiparticle dispersion is so flat in the superconducting cuprates?}
\author{Tao Li and Da-Wei Yao}
\affiliation{Department of Physics, Renmin University of China, Beijing 100872, P.R.China}
\date{\today}

\begin{abstract}
The emergence of the coherent quasiparticle peak and the development of the peak-dip-hump structure in the anti-nodal region below $T_{c}$ is the most prominent non-BCS signature of the underdoped high-$T_{c}$ cuprates, in which no coherent quasiparticle can be defined in the anti-nodal region above $T_{c}$. The peak-dip-hump structure has been commonly interpreted as the result of the coupling of the electron to some Bosonic mode. However, such an electron-Boson coupling picture does not answer the question of \textit{why the quasiparticle dispersion is so flat in the anti-nodal region}, a behavior totally unexpected for Bogoliubov quasiparticle in a d-wave BCS superconductor. Here we show that the sharp quasiparticle peak in the anti-nodal region should be understood as a new pole in the electron Green's function generated by the strong coupling of the electron to diffusive spin fluctuation around the antiferromagnetic wave vector $\mathbf{Q}=(\pi,\pi)$, rather than a nearly free Bogoliubov quasiparticle in a d-wave BCS superconductor. More specifically, we find that the normal self-energy of the electron from the scattering with the diffusive spin fluctuation manifests itself mainly as a level repulsion effect and is responsible for the reduction of both the quasiparticle dispersion and the quasiparticle dissipation rate in the anti-nodal region. We argue that the peak-dip separation in the anti-nodal spectrum should not be interpreted as the energy of the pairing glue.
\end{abstract}


\maketitle

The emergence of sharp quasiparticle peak in the anti-nodal region below $T_{c}$  is the most prominent non-BCS signature of the underdoped high-$T_{c}$ cuprates, in which no coherent quasiparticle can be defined in the anti-nodal region in the normal state\cite{ARPES1,ARPES2,ARPES3,ARPES4,ARPES5,ARPES6,ARPES7,ARPES8}. ARPES measurement shows that the coherent quasiparticle peak is almost non-dispersive in the anti-nodal region and is separated from a high energy hump structure by a dip\cite{bilayer}.  Such a peak-dip-hump structure, which has also been observed in STM measurements of the underdoped cuprates\cite{STM1,STM2,STM3}, is generally believed to be induced by the coupling of the electron to some Bosonic mode. 

The $B_{1g}$ oxygen buckling phonon mode and the $(\pi,\pi)$ spin resonance mode are the two most extensively discussed candidates for the Bosonic mode. In the electron-Boson coupling picture, the coherent peak is interpreted as the usual Bogoliubov quasiparticle in a d-wave BCS superconductor, while the hump is interpreted as the scattered quasiparticle shifted in energy by the Bosonic mode.  While such a picture does receive some experimental support, a fully consistent understanding on the origin of the peak-dip-hump phenomena is missing. In particular, it is not understood \textit{why the anti-nodal quasiparticle dispersion is so flat}, a behavior that is totally unexpected for the Bogoliubov quasiparticle in a d-wave BCS superconductor\cite{ARPES2,ARPES3,ARPES4,ARPES5,ARPES6,ARPES8}. Another strange thing about the quasiparticle peak is that it stays sharp even when the momentum is away from the underlying Fermi surface. In fact, the quasiparticle dissipation rate seems to be suppressed in the whole anti-nodal region\cite{Hashimoto}.

Here we present an alternative interpretation of the peak-dip-hump structure in the framework of the spin-Fermion model. In our theory, the coherent quasiparticle peak appear as a new pole in the electron Green's function. It is generated by the strong coupling of the electron with the diffusive spin fluctuation in the system around $\mathbf{Q}=(\pi,\pi)$, rather than with a particular Bosonic mode. The dispersion of the new pole becomes flat in the anti-nodal region as a result of a level repulsion effect related to the normal self-energy correction from such a coupling. The same level repulsion effect also pushes the quasiparticle peak to lower binding energy, resulting in a suppressed quasiparticle dissipation rate in the whole anti-nodal region.  

The antiferromagnetic spin fluctuation is widely believed to be the major pairing glue in the high-$T_{c}$ cupartes. The persistence of the antiferromagnetic spin fluctuation deep in the paramagnetic phase of the doped cuprates has been confirmed by extensive NMR and neutron scattering studies in the early days of high-$T_{c}$ study\cite{MMP,Zha,Lee}. More recently, RIXS measurements find that the high energy spin-wave-like fluctuation is robust against doping even in the overdoped regime\cite{RIXS1,RIXS2,RIXS3,RIXS4,RIXS5,RIXS6}, with its dispersion and spectral intensity only weakly dependent on doping. 

The dual nature of the electron as both itinerant quasiparticle and local moment poses a serious challenge to the theory of the high-$T_{c}$ cupates. However, at a phenomenological level, one can treat these two kinds of electron motion as independent degree of freedom and assume a phenomenological coupling between the two. The result is the so called spin-Fermion model\cite{NAFL1,NAFL2,NAFL3,NAFL4,NAFL5}, in which the local moment is assigned phenomenologically with a nearly critical dynamics, while the action of the itinerant quasiparticle is determined by fitting the ARPES results. At a microscopic level, one can understand such a separation of the electron degree of freedom in the renormalization group sense, in which it is understood that the electron at the higher energy scale behaves more like a local moment and that the electron around the Fermi energy behaves more like an itinerant quasiparticle.

The spin-Fermion model has been extensively used in the study of the high-$T_{c}$ cuprates\cite{NAFL1,NAFL2,NAFL3,NAFL4,NAFL5}. In particular, the theory provides a natural understanding on the origin of the d-wave pairing in the superconducting state\cite{NAFL1,NAFL2,NAFL5}.  The scattering from the antiferromagnetic spin fluctuation is also widely believed to be responsible for the large electron scattering rate and the pseudogap phenomena in the normal state\cite{NAFL3,NAFL4}. We thus expect that it can also have important effect on the quasiparticle dynamics in the superconducting state.

Adopting the spin-Fermion model, we show that the quasiparticle dynamics in the anti-nodal region is indeed strongly renormalized by the scattering from the diffusive spin fluctuation around $\mathbf{Q}=(\pi,\pi)$. More specifically, we find that the normal part of the electron self-energy manifests itself mainly as a level repulsion effect and is responsible for the reduction of both the quasiparticle dispersion and the quasiparticle dissipation rate in the anti-nodal region. On the other hand, the anomalous self-energy correction acts mainly to renormalize the pairing potential and can be absorbed as a redefinition of the latter. We argue that the sharp quasiparticle peak observed in the anti-nodal region should be understood as a strongly renormalized new pole in the electron Green's function, rather than a nearly free Bogoliubov quasiparticle in a d-wave BCS superconductor.

The spin-Fermion model takes the form of $H=H_{\mathrm{BCS}}+H_{int}$ in the superconducting state, in which $H_{\mathrm{BCS}}$ is the mean field Hamiltonian of a BCS superconductor and is given by
 \begin{equation}
 H_{\mathrm{BCS}}=\sum_{\mathbf{k},\sigma}\epsilon_{\mathbf{k}}c^{\dagger}_{\mathbf{k},\sigma}c_{\mathbf{k},\sigma}+\sum_{\mathbf{k}}\Delta_{\mathbf{k}}(c^{\dagger}_{\mathbf{k},\uparrow}c^{\dagger}_{-\mathbf{k},\downarrow}+h.c.).
 \end{equation}
Here $\epsilon_{\mathbf{k}}=-2t(\cos k_{x}+\cos k_{y})-4t'\cos k_{x}\cos k_{y}-\mu$ is the bare dispersion of the quasiparticle, $\Delta_{\mathbf{k}}=\frac{\Delta}{2}(\cos k_{x}-\cos k_{y})$ is a d-wave pairing potential. In this study, we set $t=250 \ meV$, $t'=-0.3t$, $\mu=-t$ and $\Delta=30 \ meV$. The doping level is thus about $x=0.145$. 
$H_{int}$ is the interaction between the itinerant quasiparticle and the local moment. It is given by
\begin{equation}
H_{int}=g\sum_{i}\vec{\mathbf{S}}_{i}\cdot\vec{\mathbf{s}}_{i},
\end{equation} 
in which $\vec{\mathbf{s}}_{i}=\frac{1}{2}\sum_{\alpha,\beta}c^{\dagger}_{i,\alpha}\vec{\mathrm{\sigma}}_{\alpha,\beta}c_{i,\beta}$ is the spin density operator of the itinerant electrons at site $i$ and $\vec{\mathbf{{S}}}_{i}$ is the local moment operator on the same site. $g$ is a phenomenological coupling constant. 

In this paper, we will adopt the widely used Monien-Mills-Pines(MMP) form\cite{MMP} as a phenomenological guess of the dynamical spin susceptibility of the local moment, which is given by
 \begin{equation}
\chi(\mathbf{q},\omega)=\frac{\chi_{0}}{1+(\mathbf{q}-\mathbf{Q})^{2}\xi^{2}-i\omega/\omega_{sf}}.
 \end{equation}
 Here $\chi_{0}\propto \xi^{2}$ is the static spin susceptibility at the antiferromagnetic wave vector $\mathbf{Q}=(\pi,\pi)$, $\xi$ is the spin correlation length, $\omega_{sf}$ is a phenomenological parameter describing the dissipation of the local moment caused by its coupling to the itinerant quasiparticles.  In this study, we set $\xi=3 a$ and $\omega_{sf}=15  \ meV$, which are typical values of these parameters for the underdoped cuprates\cite{Zha}. In the superconducting state, we should expect the low energy end of the spin fluctuation spectrum to be modified to accommodate both the spin gap and the $(\pi,\pi)$ spin resonance mode. Here we simply remove the spin fluctuation spectral weight below a momentum dependent spin gap $\Delta^{s}_{\mathbf{q}}=\mathrm{Min}_{\mathbf{k}}(E_{\mathbf{k}}+E_{\mathbf{k+q}})$, in which $E_{\mathbf{k}}=\sqrt{\epsilon^{2}_{\mathbf{k}}+\Delta^{2}_{\mathbf{k}}}$ is the mean field excitation energy of the Bogoliubov quasiparticle. The spin resonance mode is discarded in our study, since its spectral weight is too small as compared to the total spectral weight contained in $\chi(\mathbf{q},\omega)$ to generate any significant contribution to the electron self-energy\cite{Kivelson}. Lastly, we note that the integrated spectral weight of the MMP susceptibility actually diverges logarithmically at high energy. To remove such an unphysical divergence, we cut off the spectum at $\omega_{c}=30  \omega_{sf}=450 \ meV$. Such a choice of $\omega_{c}$ is consistent with the RIXS measurement on high-$T_{c}$ cuprates\cite{RIXS1,RIXS2,RIXS3,RIXS4,RIXS5,RIXS6}. 
 
We will adopt the Nambu formalism and define the electron Green's function as follows
\begin{equation}
G(\mathbf{k},\tau)=-<T_{\tau}\psi_{\mathbf{k}}(\tau)\psi^{\dagger}_{\mathbf{k}}(0)>,
\end{equation} 
in which $\psi^{\dagger}_{\mathbf{k}}=(c^{\dagger}_{\mathbf{k},\uparrow},c_{-\mathbf{k},\downarrow})$ is the Nambu spinor. If we treat the coupling between the itinerant electron and local moment perturbatively, then to the lowest order in perturbation theory the electron self-energy is given by
\begin{equation}
\Sigma(\mathbf{k},i\nu)=3\times\frac{g^{2}}{4N\beta}\sum_{\mathbf{q},i\omega}\chi(\mathbf{q},i\omega)G^{0}(\mathbf{k-q},i\nu-i\omega).
\end{equation}
Here the factor 3 comes from the fact that the three components of the local moment fluctuate with equal amplitude in the paramagnetic phase. $G^{0}(\mathbf{k},i\nu)=(i\nu -\epsilon_{\mathbf{k}}\tau_{3}-\Delta_{\mathbf{k}}\tau_{1})^{-1}$ is the Green's function of the BCS mean field state, in which $\tau_{1}$
 and $\tau_{3}$ are the usual Pauli matrix in the Nambu space. Similarly, we can decompose the electron self-energy as 
 \begin{equation}
 \Sigma(\mathbf{k},i\nu)=\Sigma^{(0)}(\mathbf{k},i\nu) +\Sigma^{(1)}(\mathbf{k},i\nu)\tau_{3}+\Sigma^{(2)}(\mathbf{k},i\nu)\tau_{1}.
 \end{equation}
 Here $\Sigma^{(0)}(\mathbf{k},i\nu)$ and $\Sigma^{(1)}(\mathbf{k},i\nu)$ are the normal part of the electron self-energy, $\Sigma^{(2)}(\mathbf{k},i\nu)$ is the anomalous part of the electron self-energy. 
  
 Using spectral representation, the electron self-energy can be expressed as
 \begin{equation}
 \Sigma^{(r)}(\mathbf{k},i\nu)=\frac{1}{2\pi}\int d\omega \frac{-2\mathrm{Im} \Sigma^{(r)}(\mathbf{k},\omega)}{i\nu-\omega},
 \end{equation}
in which $r=0,1,2$. $-2\mathrm{Im}\Sigma^{(r)}(\mathbf{k},\omega)$ is the spectral function of the self-energy in the $r$-th channel and is given by
\begin{eqnarray}
-\mathrm{Im}\Sigma^{(0)}(\mathbf{k},\omega)=&\frac{c}{N}&\sum_{\mathbf{q},s=\pm1}[n_{B}(\omega+sE_{\mathbf{k-q}})+f(sE_{\mathbf{k-q}})]\nonumber \\
&\times& R(\mathrm{q},\omega+sE_{\mathbf{k-q}})\nonumber\\
-\mathrm{Im}\Sigma^{(1)}(\mathbf{k},\omega)=-&\frac{c}{N}&\sum_{\mathbf{q},s=\pm1}[n_{B}(\omega+sE_{\mathbf{k-q}})+f(sE_{\mathbf{k-q}})]\nonumber \\
&\times& \frac{s\epsilon_{\mathbf{k-q}}}{E_{\mathbf{k-q}}}R(\mathbf{q},\omega+sE_{\mathbf{k-q}})\nonumber\\
-\mathrm{Im}\Sigma^{(2)}(\mathbf{k},\omega)=-&\frac{c}{N}&\sum_{\mathbf{q},s=\pm1}[n_{B}(\omega+sE_{\mathbf{k-q}})+f(sE_{\mathbf{k-q}})]\nonumber \\
&\times& \frac{s\Delta_{\mathbf{k-q}}}{E_{\mathbf{k-q}}}R(\mathbf{q},\omega+sE_{\mathbf{k-q}}),
\end{eqnarray}
in which $c=3g^{2}/32\pi$. $n_{B}(x)$ and $f(x)$ are the Bose and Fermi distribution function. $R(\mathbf{q},\omega)=-2\mathrm{Im}\chi(\mathbf{q},\omega)$ is the spectral function of the local spin fluctuation. 
We note that $\Sigma^{(2)}$ given above is actually an overestimation when there is either thermal or quantum phase fluctuation in the pairing potential $\Delta_{\mathbf{k}}$. To solve this problem, we introduce a Debye-Waller fact $\alpha$ for $\Delta_{\mathbf{k}}$ and renormalize the anomalous self-energy down to $\alpha\Sigma^{(2)}$.

\begin{figure}
\includegraphics[width=9cm]{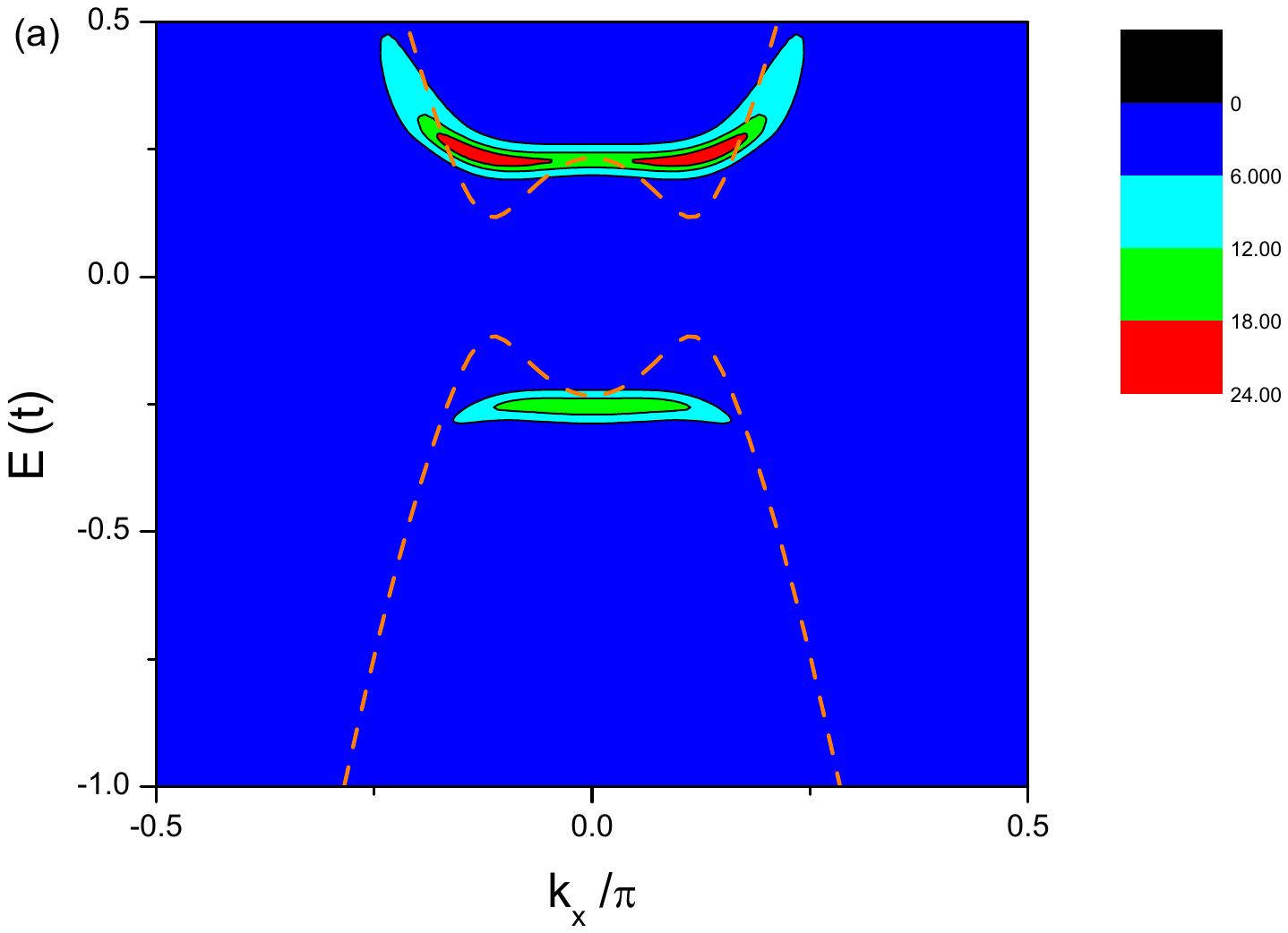}
\includegraphics[width=9cm]{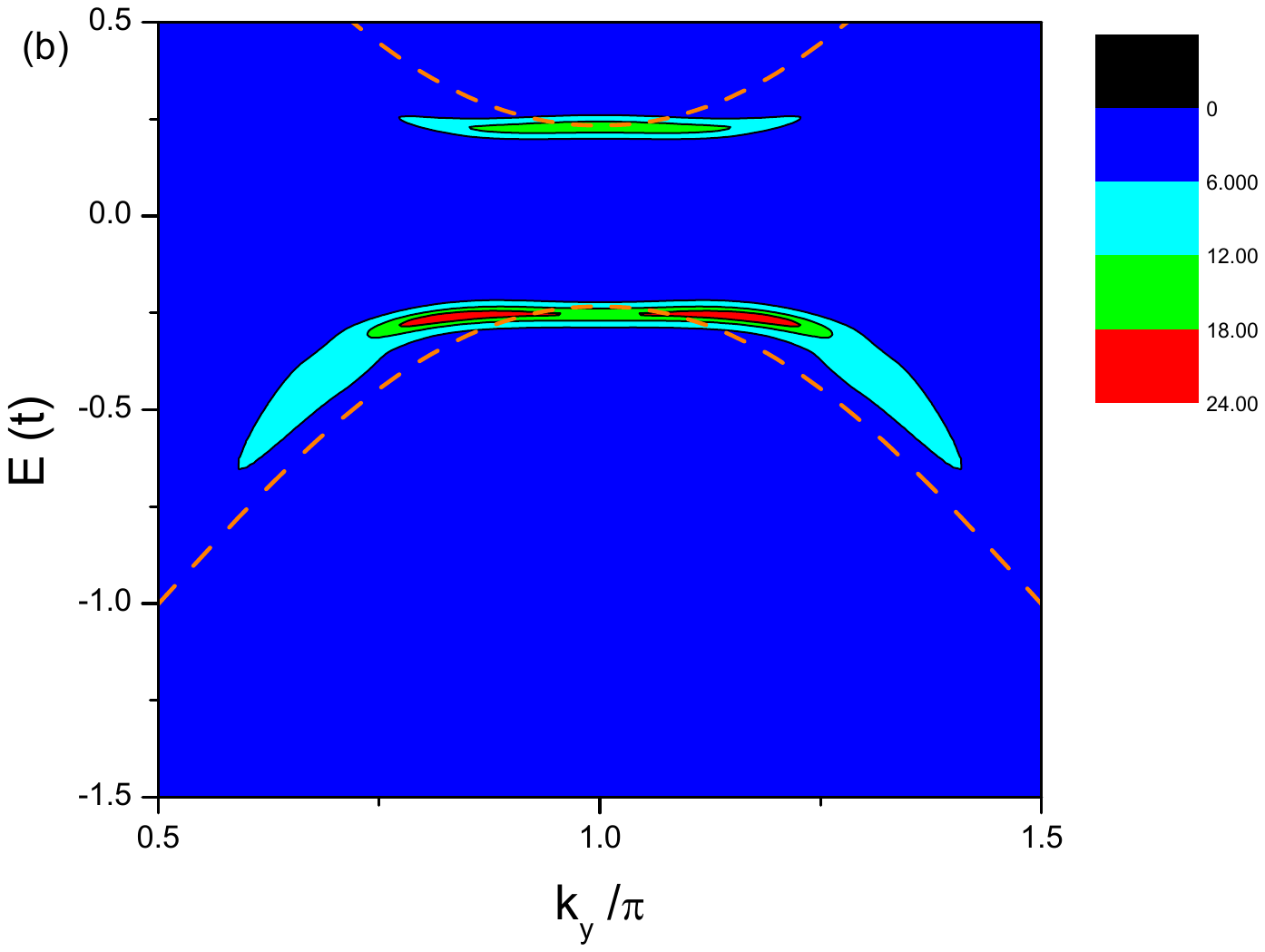}
\caption{\label{fig1}
(Color on-line) The electron spectral function around the anti-nodal point $\mathbf{M}=(0,\pi)$ along (a)the horizontal direction($\mathbf{k}=(k_{x},\pi)$) and (b)the vertical direction ($\mathbf{k}=(0,k_{y})$). The dashed lines indicate the bare BCS dispersion. In this calculation, the Debye-Waller factor $\alpha$ is set to be $1$ and a broadening of $\delta=7.5 \ meV$ is used.}
\end{figure}

The dressed electron Green's function is given by 
\begin{eqnarray}
G(\mathbf{k},i\nu)=\frac{1}{i\nu-\Sigma^{(0)}-(\epsilon_{\mathbf{k}}+\Sigma^{(1)})\tau_{3}-(\Delta_{\mathbf{k}}+\Sigma^{(2)})\tau_{1})}.\nonumber\\
\end{eqnarray}
The corresponding electron spectral function is given by 
\begin{eqnarray}
&&A(\mathbf{k},\omega)=-2 \mathrm{Im} G_{1,1}(\mathbf{k},\omega+i\delta)\nonumber\\
&=&-2\mathrm{Im}\frac{\omega+i\delta-\Sigma^{(0)}+\epsilon_{\mathbf{k}}+\Sigma^{(1)}}{(\omega+i\delta-\Sigma^{(0)})^{2}+(\epsilon_{\mathbf{k}}+\Sigma^{(1)})^{2}+(\Delta_{\mathbf{k}}+\Sigma^{(2)})^{2}},\nonumber\\
\end{eqnarray}
in which the self-energy $\Sigma^{(r)}$ should be calculated at frequency $\omega+i\delta$.  Here $\delta$ is a infinitesimal broadening of the electron spectral function.

In our study, we will focus on the electron spectral function at zero temperature. The only free parameter in the theory is then the product $g^{2}\chi_{0}$. The value of $\chi_{0}$ can be estimated by requiring that the total spin fluctuation spectral weight to satisfy the following local spin sum rule
\begin{equation}
<\mathbf{S}^{2}_{i}>=\frac{1}{(2\pi)^{3}}\int^{\omega_{c}}_{0} d\omega d\mathbf{q} R(\mathbf{q},\omega)=\frac{3}{4}(1-x).
\end{equation}
Inserting the value of $\omega_{c}$, $\omega_{sf}$ and $\xi$ and complete the integral, one find that for $x=0.145$ we have $\chi_{0}\simeq400 \ eV^{-1}$, a value which is close to the estimation from the NMR data\cite{Zha}.
In our study, we set $g=6t=1.5 \ eV$. We find that such a coupling strength is strong enough to generate significant renormalization of the quasiparticle dynamics in the anti-nodal region.

\begin{figure}
\includegraphics[width=9cm]{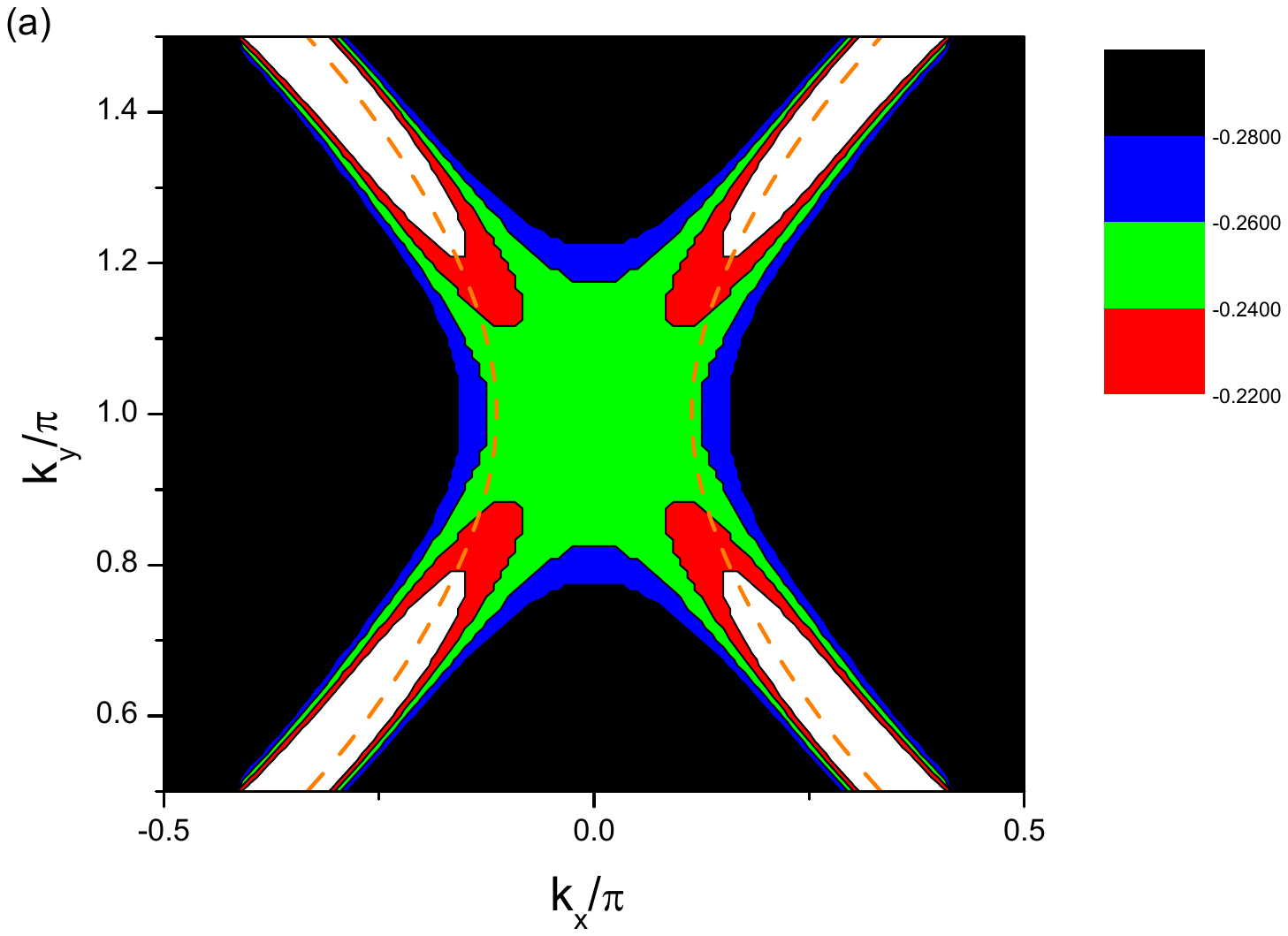}
\includegraphics[width=9cm]{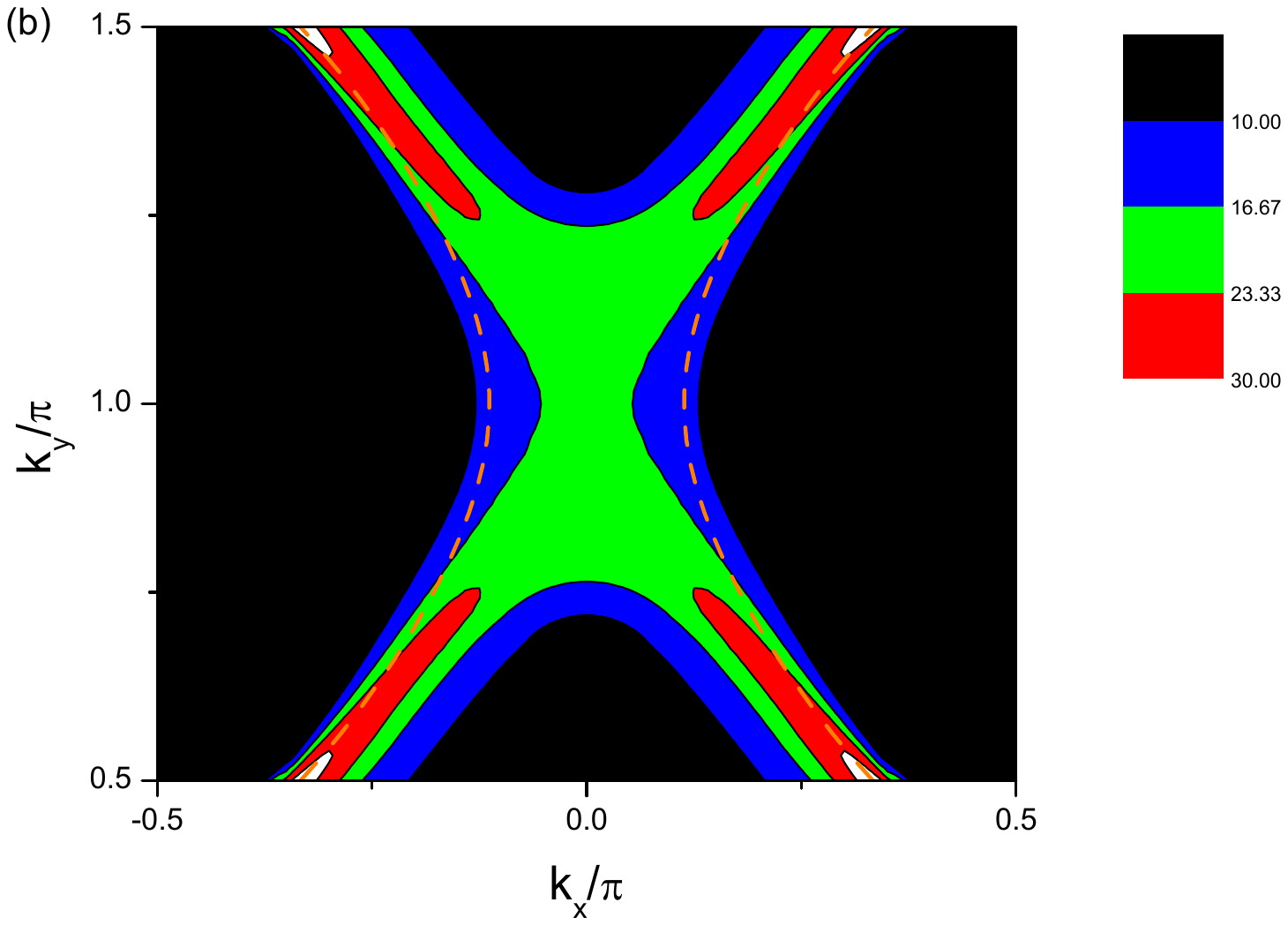}
\caption{\label{fig2}
(Color on-line) The energy (a) and the peak intensity(b) of the quasiparticle peak around the anti-nodal point $\mathbf{M}=(0,\pi)$. Here we use $t$ as the unit of energy and the height of the quasiparticle peak as a relative measure of the peak intensity. The dashed lines in the plots mark the position of the underlying Fermi surface.In this calculation, the Debye-Waller factor $\alpha$ is set to be $1$ and a broadening of $\delta=7.5 \ meV$ is used}
\end{figure}

The electron spectral function around the anti-nodal point $\mathbf{M}=(0,\pi)$ is plotted in Fig.1 along a line crossing the $\mathbf{M}$ point in both the horizontal and the vertical direction. The most prominent feature of the spectral function is the sharp and flat quasiparticle band around the $\mathbf{M}$ point. In fact, we find that such a sharp and flat quasiparticle band exists in a whole two dimensional area around the $\mathbf{M}$ point(see Fig.2), in which the total dispersion is less than $5 \ meV$. This is totally unexpected from the BCS mean field theory of a d-wave superconductor. Moreover, we find that the quasiparticle peak remains sharp even when the momentum is far away from the mean field Fermi surface. Both of these features are consistent with experimental observations\cite{ARPES2,ARPES3,ARPES4,ARPES5,ARPES6,ARPES8}.

To see more clearly the spectral weight distribution in the broad continuum, which is overwhelmed by the sharp quasiparticle peak in Fig.1, we plot the electron spectral function at the $\mathbf{M}$ point in Fig.3. A characteristic peak-dip-hump structure can be clearly seen in the spectrum. We find that the sharp quasiparticle peak and the related peak-dip-hump structure persist in the whole anti-nodal region. 
\begin{figure}
\includegraphics[width=8cm]{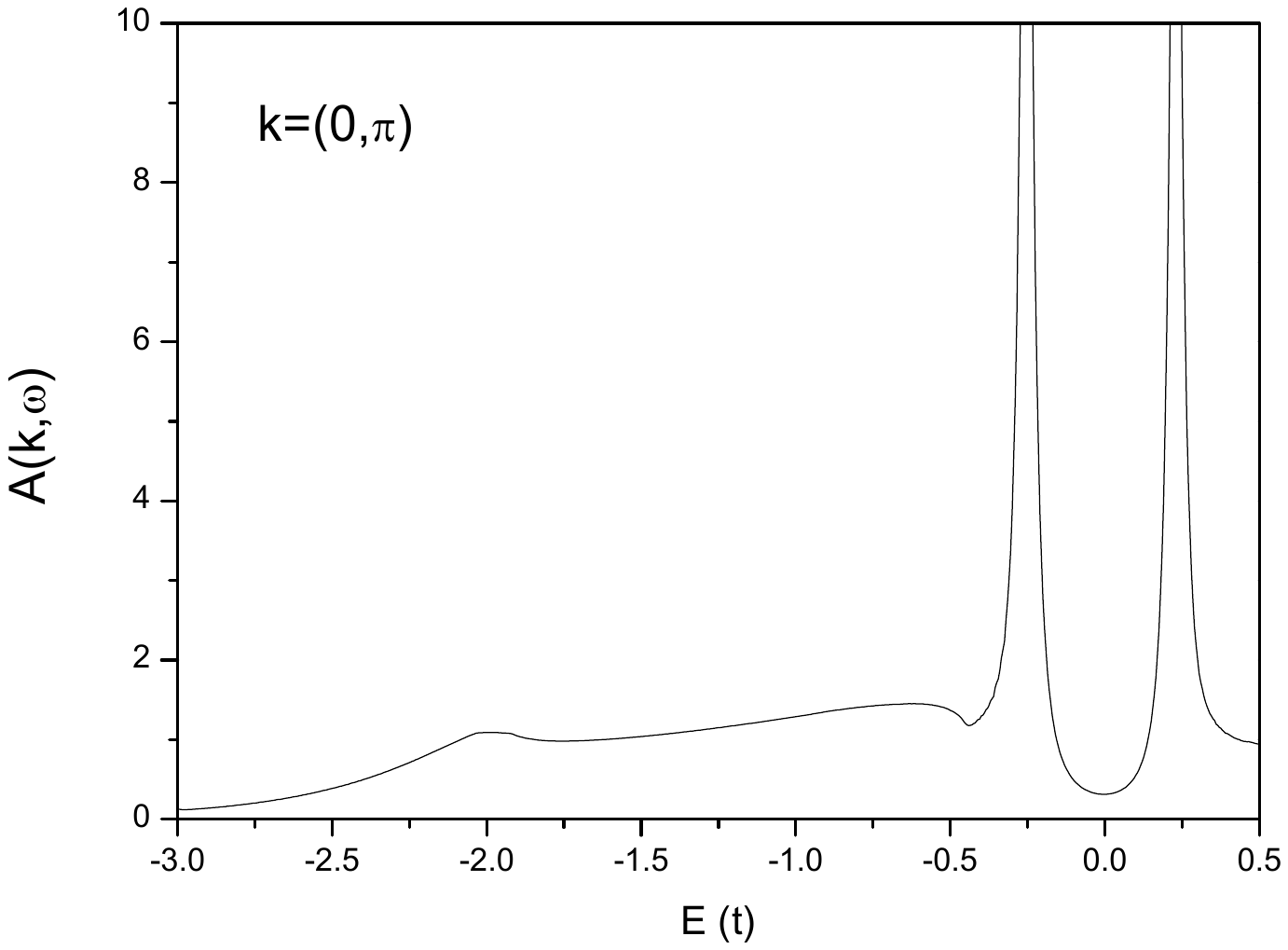}
\caption{\label{fig3}
The electron spectral function at the anti-nodal point. The sharp peak and the characteristic peak-dip-hump structure are found to persist in the whole anti-nodal region. The maximum at $-2 t$ is an artifact induced by our hard cutoff in the spin fluctuation spectrum at $\omega_{c}$.}
\end{figure}

We now analyze the origin of such a peculiar behavior of the anti-nodal quasiparticle in the spin-Fermion model. We find that the sharp quasiparticle peak in the anti-nodal region is actually a new pole in the electron Green's function and is generated by the strong coupling of the electron with the diffusive spin fluctuations around $\mathbf{Q}=(\pi,\pi)$. We find that the normal and the anomalous self-energy from such a coupling play very different roles in the quasiparticle dynamics.

\begin{figure}
\includegraphics[width=8cm]{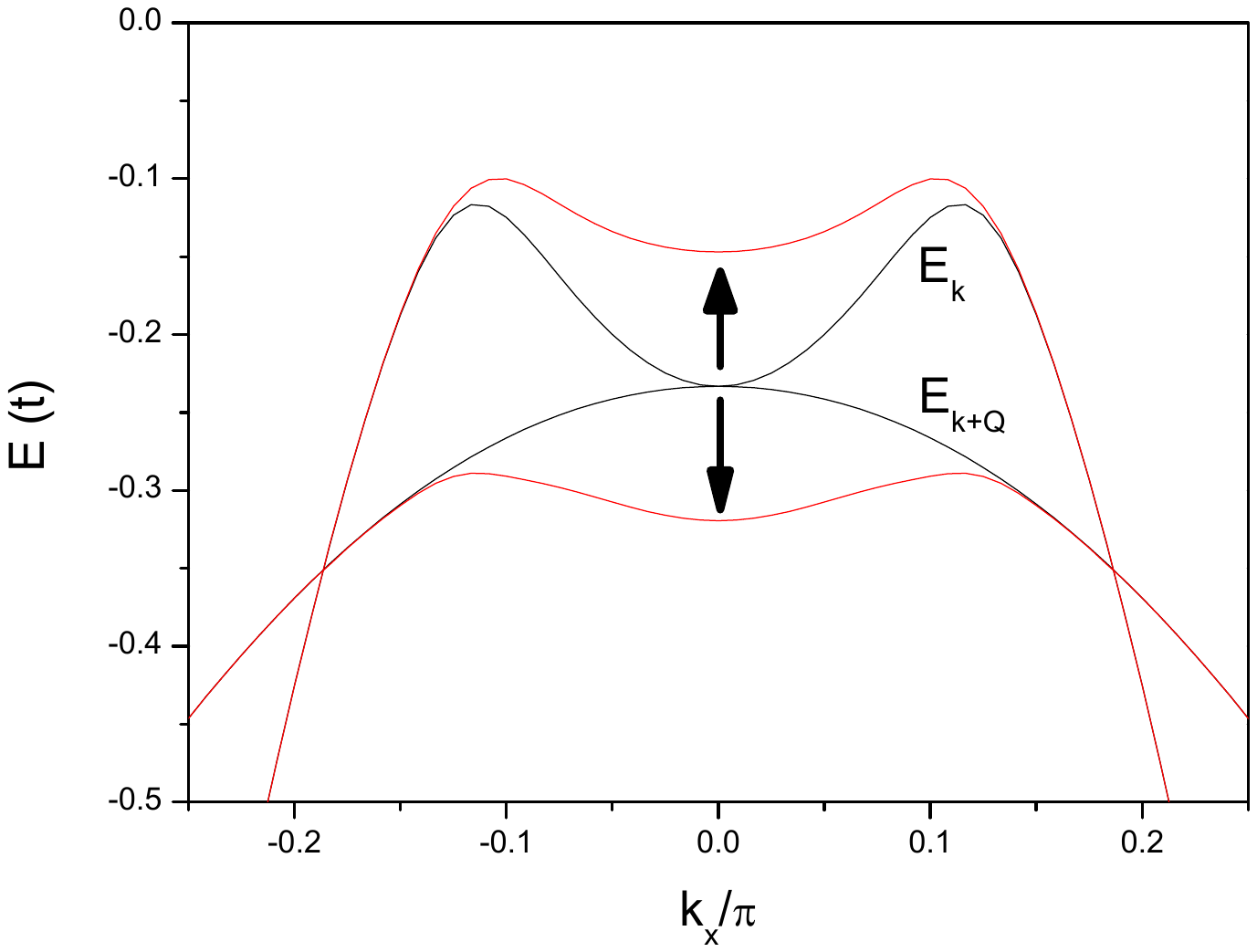}
\caption{\label{fig5}
 Illustration of the level repulsion effect between the quasiparticle states around the anti-nodal point. The momentum is along the horizontal direction, namely, $\mathbf{k}=(k_{x},\pi)$ in this plot. The black lines denote the unperturbed levels at $E_{\mathbf{k}}$ and $E_{\mathbf{k+Q}}$ in the occupied side of the electron spectrum. The red lines denote the renormalized levels after we introduce a momentum independent coupling between the two unperturbed levels. We note that in the superconducting state the matrix element of the coupling Hamiltonian should be multiplied by a factor of $v^{2}_{\mathbf{k}}v^{2}_{\mathbf{k+Q}}$ on the occupied side of the electron spectrum. Since such a factor decreases rapidly when $\epsilon_{\mathbf{k}}$ exceeds zero, we expect the flat quasiparticle band to terminate approximately at $k_{\mathrm{F}}$ along the $(0,\pi)$-$(\pi,\pi)$ direction.}
\end{figure}

More specifically, we find that the normal self-energy manifests itself mainly as a level repulsion effect between the quasiparticle level $E_{\mathbf{k}}$ and the scattering final state level $E_{\mathbf{k+q}}+\Omega_{\mathbf{q}}$. Here $\Omega_{\mathbf{q}}$ denotes the energy of the spin fluctuation at momentum $\mathbf{q}$ and is distributed in a broad energy range. In Fig.4, we plot the dispersion of $E_{\mathbf{k}}$ and $E_{\mathbf{k+q}}$ around the $\mathbf{M}$ point for $\mathbf{q=Q}$, where the local spin fluctuation is the strongest. The difference between $E_{\mathbf{k}}$ and $E_{\mathbf{k+Q}}$ is found to reach its minimum at the $\mathbf{M}$ point, where we expect the strongest level repulsion effect to occur. The quasiparticle pole in the upper branch will thus be pushed to lower binding energy, resulting in a reduction in its dispersion and dissipation rate around the $\mathbf{M}$ point. This explains the emergence of the flat and coherent quasiparticle peak in the anti-nodal region. Unlike the upper branch, the spectral weight corresponding to the lower branch in Fig.4 is distributed in a broad energy range as a result of the dispersion in $E_{\mathrm{k+q}}$ and the diffusion in $\Omega_{\mathrm{q}}$. The sharp quasiparticle peak in the anti-nodal spectrum will thus be accompanied by a broad hump structure .

On the other hand, the anomalous self-energy acts mainly to renormalize the bare pairing potential. This can be seen clearly in Fig.5, from which one find that the flat quasiparticle band simply shift down in binding energy when we decrease the value of the Debye-Waller factor $\alpha$. We can thus simply absorb the effect of $\Sigma^{(2)}$ into a redefinition of $\Delta$ when we discuss the dynamics of low energy quasiparticles\cite{dwave}.

\begin{figure}
\includegraphics[width=9cm]{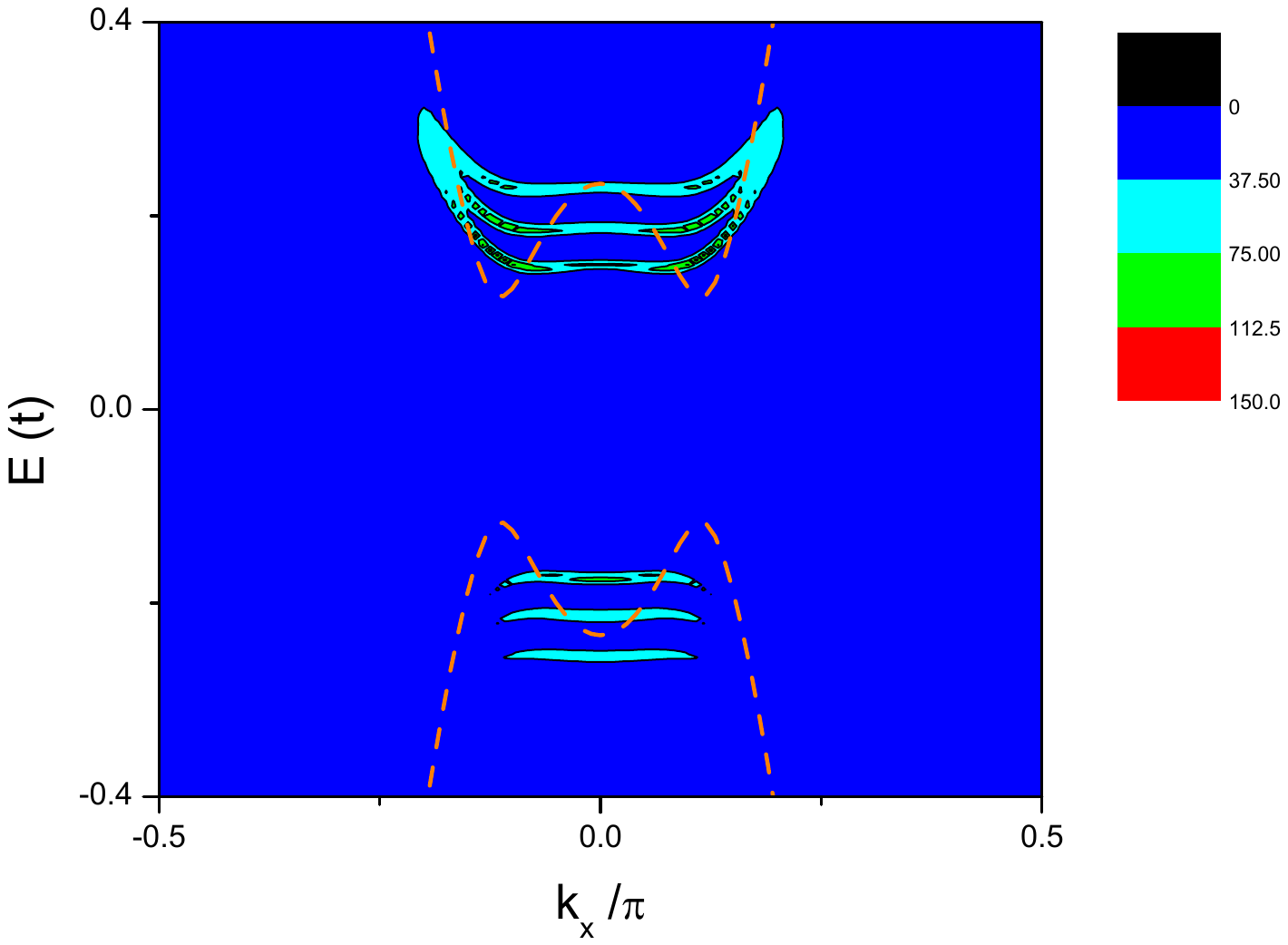}
\caption{\label{fig4}
 The electron spectral function along the horizontal direction($\mathbf{k}=(k_{x},\pi)$) for three different values of the Debye-Waller factor $\alpha=1.0$, $0.8$ and $0.6$. The quasiparticle bands move to higher binding energy with the increase of $\alpha$. The dash lines indicate the bare BCS dispersion. In this plot, a smaller broadening of $\delta=2.5 \ meV$ is used to resolve the spectral weights for different $\alpha$.}
\end{figure}

In the level repulsion picture discussed above, a momentum transfer of $\mathbf{Q}=(\pi,\pi)$ in the electron scattering process is crucial for the emergence of the flat quasiparticle band in the anti-nodal region. This argues against a phonon origin of the peak-dip-hump structure in the cupartes, since the $B_{1g}$ oxygen buckling phonon mode does not posses such a characteristic momentum. The coupling to the $(\pi,\pi)$ spin resonance mode is also not likely the origin for the flat quasiparticle band, although it has the right momentum. The reason is two fold. First, the spectral weight of the spin resonance mode is very small and is at the same time strongly temperature dependent. It is hard to believe that such a small spectral weight can produce the strong and almost temperature independent quasiparticle renormalization effect observed in the ARPES experiment. Second, the spin resonance mode has a well defined energy. It is thus hard to believe that it can produce the broad and almost featureless hump structure in the ARPES spectrum. In the spin-Fermion model, it is the diffusive fluctuation of the local moment in a broad energy range that is responsible for such a broad and almost featureless high energy hump structure.

Previous studies have commonly used the peak-dip separation in the anti-nodal spectrum as an estimate of the Boson energy in the electron-Boson coupling picture. In our theory, such an energy is determined by the size of the spin gap(or pairing gap) and the strength of the level repulsion effect, which are proportional to each other since both are self-energy correction from the spin-Fermion coupling. It is thus not surprising to find linear correlation between the mode energy defined in this way and the energy of the spin resonance mode, which is of the order of the pairing gap. However, this correlation does not imply that the peak-dip-hump structure is induced by the spin resonance mode, whose spectral weight is too small to cause any significant self-energy correction.

Finally, we discuss the validity of the perturbative treatment adopted in this study. In principle, one should treat both the normal and the anomalous self energy correction in a self-consistent way. At the same time, one should take into account of the vertex correction of the scattering by antiferromagnetic spin fluctuation, which is especially important in the anti-nodal region. Here we will leave such a more sophisticated treatment to future studies and just mention two direct expectations from it. First, the spectral maximum of the hump structure will be pushed to higher binding energy when we include higher order scattering process in the self-consistent treatment. Second, the dispersion of the shadow band at energy $E_{\mathrm{k+q}}+\Omega_{\mathrm{q}}$ will also be greatly reduced in the self-consistent treatment. We note that these two expectations are all consistent with ARPES measurements on the Bi2201 system\cite{ARPES6}.

In conclusion, in the search for an answer to the question \textit{why the anti-nodal quasiparticle dispersion is so flat?}, we realize that the anti-nodal quasiparticle peak in the underdoped cuprates should be understood as a new pole in the electron Green's function generated by the strong coupling of the electron to diffusive spin fluctuation around the antiferromagnetic wave vector $\mathbf{Q}=(\pi,\pi)$, rather than a nearly free Bogoliubov quasiparticle in a d-wave BCS superconductor. We find that the normal self-energy of the electron from the scattering with diffusive spin fluctuation manifests itself mainly as a level repulsion effect and is responsible for the reduction of both the quasiparticle dispersion and the quasiparticle dissipation rate in the anti-nodal region. We argue that the peak-dip separation in the anti-nodal spectrum should not be interpreted as the energy of the pairing glue. We think such a strong coupling picture should apply also in other unconventional superconductors driven by spin fluctuation, in which a peak-dip-hump structure is observed in ARPES or tunneling spectrum.

We acknowledge the support from the National Natural Science Foundation of China(Grant No. 11674391), the Research Funds of Renmin University of China(Grant
No.15XNLQ03), and the National Program on Key Research Project(Grant No.2016YFA0300504).


\begin{thebibliography}{99}
\bibitem{bilayer}We note that such a nearly non-dispersive quasiparticle peak is observed in both the single-layer\cite{ARPES6} and the multi-layer cuprates\cite{ARPES2,ARPES3,ARPES4,ARPES5,ARPES8}. In the single-layer cuprates, the peak is not as prominent and manifests itself as a non-dispersive shoulder in the electron spectrum. We think such a difference in the quasiparticle peak in the single and multi-layer cuprates can be attributed to the difference in the pairing gap in these systems.
\bibitem{dwave}In principle, the anomalous self-energy can change the functional form of the pairing potential, making it deviate from the standard d-wave form in the anti-nodal region, if we have a longer spin correlation length $\xi$ or a larger coupling strength $g$.
\end{thebibliography}
\end{document}